\documentclass[showpacs,showkeys,superscriptaddress,aps,amsmath,amssymb]{revtex4}

\usepackage{t1enc}
\usepackage{graphicx}
\usepackage{bm}

\newcommand{\cis}{C_{\scriptscriptstyle IS}(t)}
\newcommand{\cisrem}{C_{\scriptscriptstyle IS}^{\rm rem}(t)}

\newcommand{\tmct}{T_{\scriptscriptstyle MCT}}
    
\begin{document}
 
\title{Inherent structures dynamics in glasses: a comparative study}
\author{D.A. Stariolo} 
\email{stariolo@if.ufrgs.br} 
\homepage{http://www.if.ufrgs.br/~stariolo}
\affiliation{Departamento de F\'{\i}sica, Universidade Federal 
	               do Rio Grande do Sul, \\
       CP 15051, 91501-970 Porto Alegre, Brazil}    
\altaffiliation{Research Associate of the Abdus Salam International
Centre for Theoretical Physics, Trieste, Italy}
\author{J.J. Arenzon} 
\email{arenzon@if.ufrgs.br} 
\homepage{http://www.if.ufrgs.br/~arenzon}
\affiliation{Departamento de F\'{\i}sica, Universidade Federal 
	               do Rio Grande do Sul, \\
       CP 15051, 91501-970 Porto Alegre, Brazil}    
\altaffiliation{Research Associate of the Abdus Salam International
Centre for Theoretical Physics, Trieste, Italy}
\author{G. Fabricius} 
\email{fabriciu@fisica.unlp.edu.ar} 
\affiliation{Departamento de F\'{\i}sica, 
Universidad Nacional de La Plata,
         \\ cc 67, 1900
	La Plata, Argentina.}
\altaffiliation{Research Associate of the Abdus Salam International
Centre for Theoretical Physics, Trieste, Italy}

\date{\today}

\begin{abstract}
A comparative study of the dynamics of inherent structures at low 
temperatures is performed on different models of glass formers: a three dimensional 
Lennard-Jones binary mixture (LJBM), facilitated spin models (either 
symmetrically constrained, SCIC, or asymmetrically, ACIC) and the trap 
model. We use suitable
correlation functions introduced in a previous work
which allow to distinguish the behaviour between models
with or without spatial or topological structure. Furthermore, 
the correlations between inherent structures behave
differently in the cases of strong (SCIC) and fragile (ACIC, LJ) glasses as a
consequence of the different role played by energy barriers when the temperature is
lowered.
 The similarities 
in the behavior of the ACIC and LJBM suggest a common nature of the 
glassy dynamics for both systems. 

\end{abstract}

\maketitle

\section{Introduction}

As the temperature of a supercooled liquid is lowered the dynamics of relaxation slows
down dramatically, leading eventually to an effective breakdown of ergodicity. The
increasing inability of the system to sample the whole phase space in experimental
timescales can be traced back by looking at the evolution in phase space, or
equivalently, by observing the mobility of particles in real space. Slow dynamics and
the difficulty to become ergodic as the temperature is lowered implies, 
for a glass former, an increasing confinement in configuration space or the reduction
in the mobility of individual particles. From a topographic view of the glass
behavior, at low temperatures a glass former will evolve
in a rough landscape and its dynamics will be influenced by the presence of many points of 
local mechanical equilibrium, called inherent structures~\cite{StWe83}, among which the 
system will wander during increasing times as the temperature is decreased. 
Thus, in this picture, confinement is a natural consequence of the non trivial
structure of the landscape at low temperatures. However, confinement may be
also due to a strong reduction of allowed paths, not related to the
underlying potential energy landscape (PEL) but, instead, to purely dynamical 
constraints. 
Nevertheless, a non trivial landscape is not necessary in order to observe other glassy
features. For example, in a model of traps with a random distribution of 
energies in an
otherwise flat landscape, dynamics proceeds exclusively through activation over energy
barriers, and it shows several glassy features, like non-exponential relaxation, aging
and a glass like transition \cite{MoBo96}. Nevertheless, by construction it is clear that no
confinement in configuration space is possible in this model.

The introduction of inherent structures (IS) (local minima of the 
PEL) \cite{StWe83}, which divide 
 the phase space into basins of attraction, allows the 
 separation of vibrational motion from the more fundamental structural
 transitions. Activation over barriers between inherent
structures and the escape through saddles with many unstable directions in the 
potential energy landscape are the relevant mechanisms for relaxation from
a landscape perspective.

In a previous work \cite{FaSt03} we introduced time correlation functions between 
inherent structures that display useful information on the relaxation
properties of glassy systems, like different dynamical regimes and stretched
exponential relaxation. But more interestingly they clearly show the signature of
confinement in configuration space as temperature is lowered
and allow one to distinguish between different systems
with glassy properties.  For a Lennard-Jones binary mixture (LJBM), these correlations 
present two well separated time 
regimes, respectively corresponding to the exploration of the interior of the basin 
of a particular IS and to the neighborhood of that basin \cite{FaSt03}.
The behaviour of the 
correlations as temperature is lowered implies the emergence of a strong
confinement in configuration space, which is absent when the same 
functions are computed for a walker moving in a cubic lattice of traps 
\cite{FaSt03}. 
In this paper we compare the behavior of these correlations for several prototypical
models of glass formers and glassy behavior: the Lennard-Jones binary mixture (LJBM),
the trap model and two kinetically constrained Ising models. While all four systems
display many of the characteristic features of supercooled liquids and glasses,
clearly the physical mechanisms behind their complex behavior are very different.
The LJBM is a very much studied molecular glass former with a complex landscape
responsible for its glassy behavior at low temperatures. The trap model, on the
other hand, has a
trivial landscape, like a flat golf court with holes with a random distribution of 
depths. Both the symmetrically (SCIC) and asymmetrically (ACIC) constrained Ising 
chain \cite{RiSo03} are intermediate between the previous two 
extreme models:
while the landscape responsible for the thermodynamics is trivial, the dynamics is defined
independently of it in order to restrict severely the possible paths in phase space,
therefore producing a very interesting glassy behavior. Our results point to a common
mechanism for relaxation in the Lennard-Jones and the kinetically constrained models
while the trap model behave in a completely different way, reflecting the difference
in the structure of the effectively sampled phase space.

In section \ref{correlations} we define the time correlation functions between inherent
structures analyzed in this work. In section \ref{LJ} and \ref{traps} we 
briefly discuss the results for the BLJM and trap model 
(for details see ref.~\onlinecite{FaSt03}). In section \ref{kinetic}
we present and discuss our results for the kinetically constrained Ising chains
and finally in section \ref{final} we present
our conclusions. 

\section{Time correlation functions between Inherent Structures}
\label{correlations}

Every instantaneous state of the system may be associated with an IS of the energy
landscape. In the LJBM, for example, following a zero temperature steepest descent 
path starting at the equilibrium initial state, the final state corresponds to the 
IS. We have observed that in this case, the IS can be identified by their energy
since degeneracies (interchange of two particles of the same kind,
for example) are quite unusual and has no significant statistical weight.
On the other hand, in a spin model the inherent structures correspond to 
configurations that are stable under single spin flip dynamics. As a consequence, 
in the kinetically constrained Ising models IS correspond to configurations
with {\em isolated defects}. In order to compute the IS in these models we start from
an equilibrium configuration at temperature $T$ and turn all excitations down,
performing a quench at zero temperature until the stable configuration is
attained. At variance with the Hamiltonian LJBM, in the kinetically constrained 
models the final inherent structures reached are not unique given an initial 
equilibrium configuration. Instead the final point depends on the actual path of 
single spin flips in the zero temperature dynamics. Nevertheless we verified that 
for different single spin flip paths the final averaged quantities of interest, the 
correlations in this case,  are invariant. In the trap model, by construction,
every configuration corresponds to an IS.

Once the IS is obtained, we measure equilibrium correlation functions that 
provide information on how confined in a region of the configuration space is the
system. The first of these correlations, $\cis$, measures the probability of being in
the same IS at two different times. More precisely, given the IS 
corresponding to the configuration at $t=0$, we define $\cis$ as the probability 
that the system {\em is at the same} IS after a time $t$, irrespective of where the 
system was in between:
\begin{equation}
\cis =\frac{1}{N_t} \sum _{i=1}^{N_t} \delta _{t_i , t_i + t}
~ , ~~~~~{\rm with} ~~ \delta _{t_i , t_j}=
           \left\{
	   \begin{array}{l}
	   1 ~ {\rm if~ } {\cal M}_{IS}(t_i)={\cal M}_{IS}(t_j)\\
	   0 ~ {\rm if~  not} 
	   \end{array} \right.
\label{eqcis}
\end{equation}
where ${\cal M}_{IS}(t_i)$ is the inherent structure configuration at time $t_i$
and with the sum we perform an average over the $N_t$ available times. Another
possible measure, $\cisrem$ is the probability of staying in the same IS for all 
times between 0 and $t$, the persistence:
\begin{equation}
\cisrem =\frac{1}{N_t} \sum _{i=1}^{N_t} \prod_{t'=0}^t \delta _{t_i , t_i + t'}
\label{eqcisrem}
\end{equation}
where the productory is restricted to the interval $[0,t]$.
These correlation functions are size dependent. We showed in \cite{FaSt03} that 
in the LJBM the behavior is in agreement with the assumption that the whole system 
is divided into nearly independent subsystems with a typical length scale. 
This allows one to rationalize the results for different sizes. The existence of a 
characteristic temperature dependent length scale, associated with the size of 
dynamical heterogeneities, has been recently proposed~\cite{GaCh02,BeGa03,Be03} and 
its consequences to the dynamics of kinetically constrained models 
extensively explored.


\section{The LJBM}
\label{LJ}

\begin{figure}[ht]
\includegraphics[width=5.5cm,height=6.8cm,angle=270]{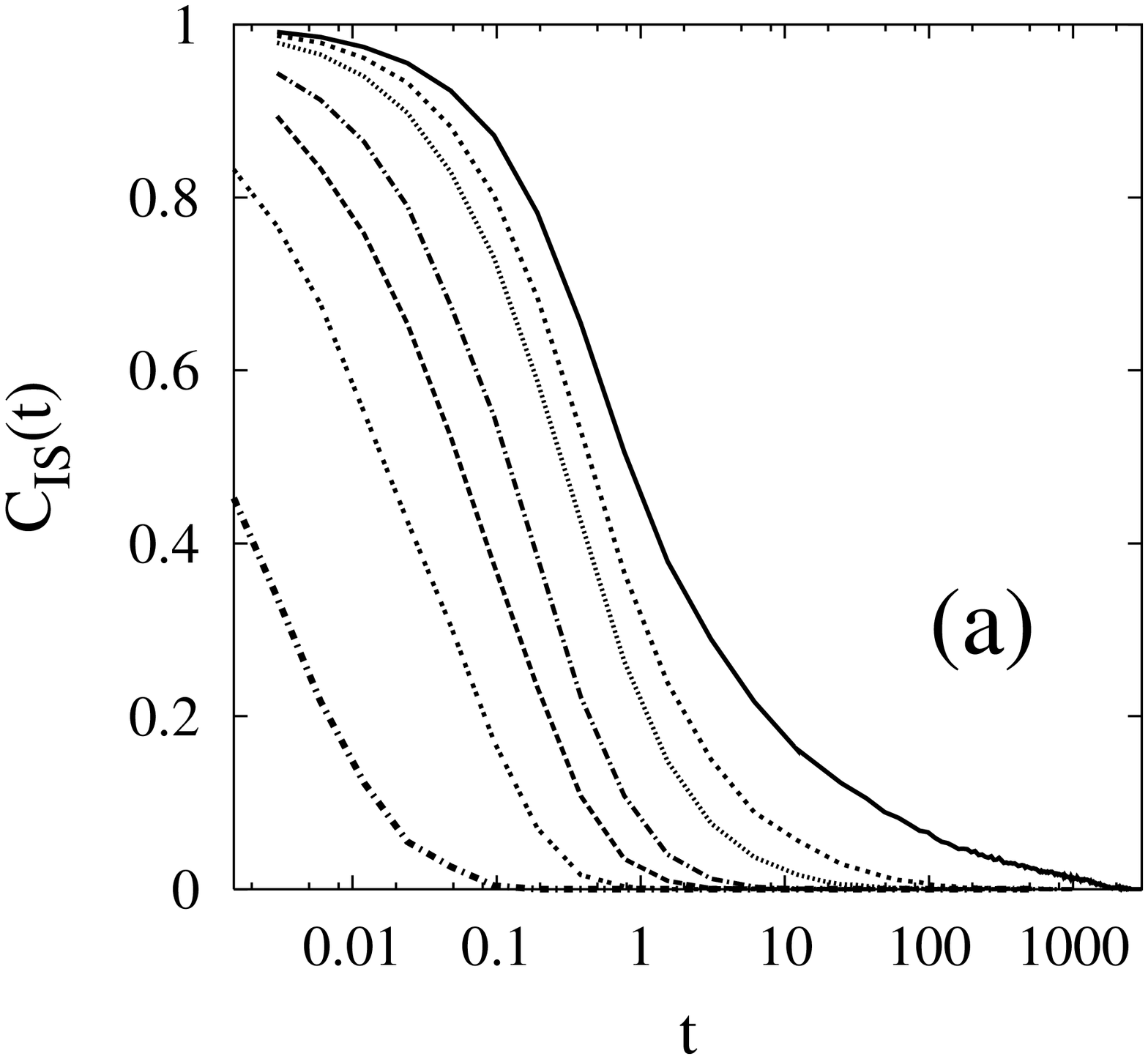}
\includegraphics[width=5.5cm,height=6.8cm,angle=270]{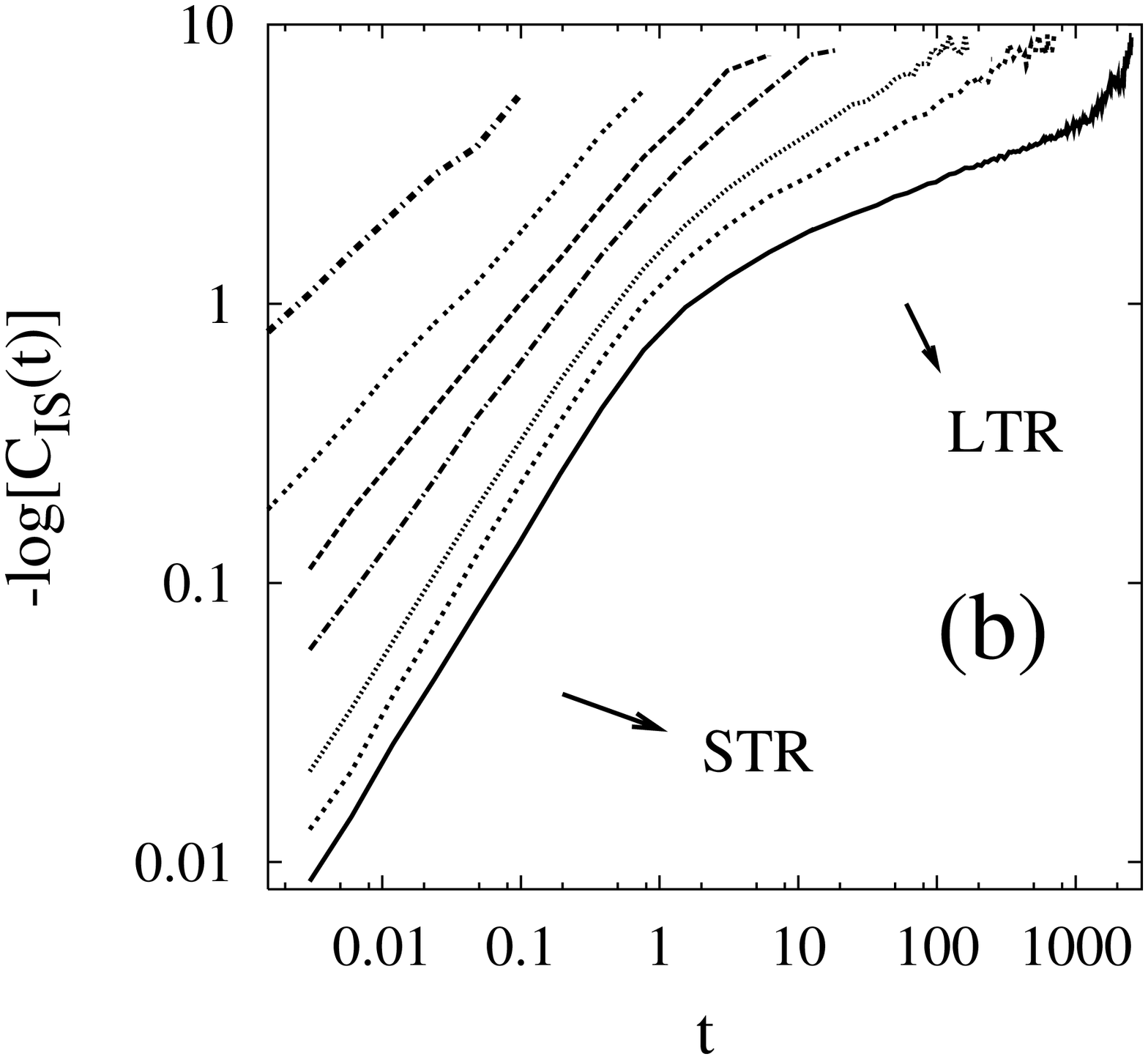}
\caption{\label{LJfig}
(a) Temperature dependence of $\cis$ for $N=130$.
From left to right, $T=2.0, 1.0, 0.8, 0.7, 0.6, 0.55$ and $0.5$.
(b) $-\log\cis$  for the same set of temperatures;
plotted in this way a stretched exponential
$exp[ -(t/\tau)^\beta$ ] is a straight line with slope $\beta$.
The short (STR) and long (LTR) time regimes are also indicated.
}
\end{figure}

Here we summarize the results of molecular dynamics simulations of a (80:20)
binary mixture Lennard-Jones system \cite{FaSt02} for $N=130$
and temperatures ranging from 0.5 to 2. In figures \ref{LJfig}a and b we 
plot $\cis$.  Figure \ref{LJfig}b shows that the low temperature behavior of
$\cis$ consists of two stretched exponentials with quite
different exponents. In the short time regime (STR) the value of $\beta$
is around $0.8$ and it is almost constant for $T$ varying between 0.6 and 0.5.
In the long time regime (LTR) the value of beta is very small ($\beta=0.2$
 for $T=0.5$) and decreases as the temperature is lowered.

\begin{figure}[ht]
\includegraphics[width=5.5cm,height=7.8cm,angle=270]{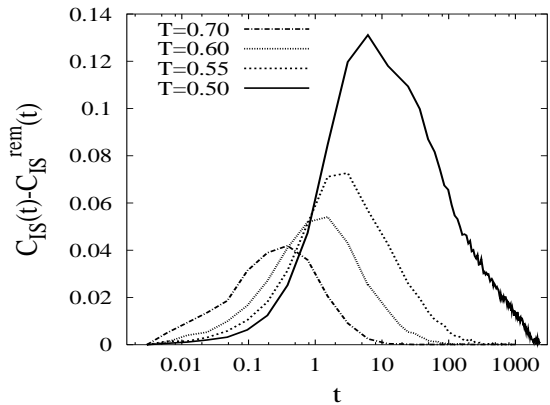}
\caption{\label{dif_LJ} The difference
$\cis-\cisrem$ for the LJ binary mixture and
different temperatures.}
\end{figure}

In order to isolate from $\cis$ the effect of remaining in a single
basin for all times, we show in  figure \ref{dif_LJ} the difference
$\cis - \cisrem$ which represents the probability of the system to be in
the same IS after a time $t$ {\em knowing that it has departed
at least once during this time interval}. 
This probability shows two important characteristics: first,  
there is a characteristic time at which the probability of returning
to the original IS is maximal. This time scale grows
when temperature is lowered: if the system goes out of a basin it takes more time
to return as the temperature is lowered. The second feature is more
important: the returning probability {\bf grows} as the glass transition
is approached. This is a signature of {\em confinement in configuration space}.

\section{The Trap Model}
\label{traps}
The model of traps in a $d$-dimensional hypercubic lattice is realized as a
random walk of a particle hopping between traps attached at each lattice
site with a given trap energy distribution $\rho(E)$ \cite{MoBo96,BoGe90}. 
Depending on this distribution, different interesting dynamical behaviors are observed.
If the energies are exponentially distributed the model has a dynamical
phase transition at a finite temperature $T_0$ below which it presents
typical glass phenomenology, such as aging effects. 
The trap model has been proposed as a
phenomenological model for describing the physics of structural
and spin glasses. From a physical point of view the dynamics proceeds 
through activation over barriers corresponding to the depth of the traps.
Besides the difference in the depths, the landscape can be considered flat,
structureless. 
Thus, in this case, we associate a trap to a basin 
and study the probability that the system returns to a given basin (trap).
We considered  two and three dimensional lattices, and assigned to 
every site a trap of energy E that is determined randomly
from an exponential distribution of the form $e^{E/T_0}$. 
The energy associated to a given
site is kept fixed during the simulation, i.e.
when the walker returns to a given site it finds
the same trap (quenched-disorder case \cite{BoGe90}).
We have used $L=100$ for $d=3$ and $L=1000$ for $d=2$. Other details of the
simulation can be found in \cite{FaSt03}.
Note that the function $\cisrem$ defined above corresponds exactly
to the equilibrium correlation $C_{eq}(t)$ defined in equation (4) 
of \cite{MoBo96} since in this model there is no difference between
the actual configuration of the system and the corresponding IS.
We verified that $\cisrem$ presents the expected long time behavior
at low temperatures: $\cisrem \sim t^{-(x-1)}$ with
$x=T/\tmct$, obtained theoretically by Monthus {\it et al}
 \cite{MoBo96} (equation (14)).

\begin{figure}[ht]
\includegraphics[width=5.5cm,height=7.8cm,angle=270]{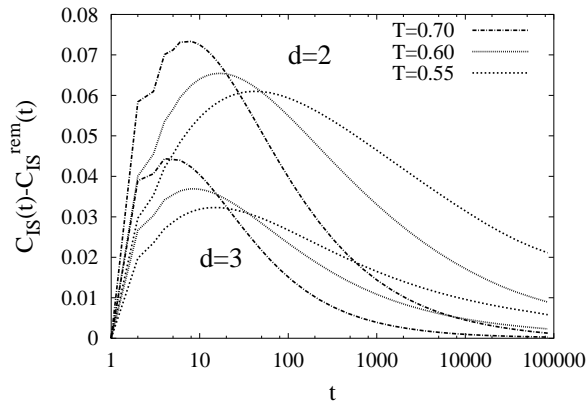}
\caption{\label{trap} The difference
$\cis-\cisrem$ for the trap model in  
$d=2$ and $d=3$ for several temperatures $T>T_0$. }
\end{figure}

In figure \ref{trap} we show the difference $\cis-\cisrem$ 
for the trap model in two and three dimensions.
The behavior is qualitatively the same in both cases. The figure shows 
a characteristic time that increases as the temperature is lowered,
analogously to figure~\ref{dif_LJ} for the LJBM.
In the case of the trap model, the maximum of the probability of
returning to a given trap moves towards increasing times as 
temperature is lowered because the walker stays for increasing times
in the traps of the surroundings. 
The confinement can only be attributed to the time spent by the walker
in the individual traps since every trap is spatially 
equivalent to each other, there is no spatial confinement
as the walker moves in a flat landscape. But the fact that distinguishes
more the trap model from the LJBM is the fact that the
peak probabilities for returning {\em decrease as the temperature is
lowered}. For lower temperatures, if the walker goes out of a trap, the
probability to come back diminishes as a direct consequence of the lack of
structure of the landscape in which it moves, which is effectively flat.

\section{Kinetic models with dynamical constraints}
\label{kinetic}

In this section we consider a class of models that, in some sense, are intermediate 
between the LJBM with its complex landscape and the trap model with its
flat landscape. Kinetic Ising models with dynamic constraints are defined as a
set of spins $n_i=0,1$ on a $d$-dimensional lattice without an explicit interaction
Hamiltonian \cite{RiSo03}. Complex behavior arises through the dynamics defined in such 
a way that a spin can flip only if it satisfies a constraint imposed on the 
number of nearest neighbors up spins. The dynamic rules are chosen  such 
that the equilibrium distribution corresponds to that of a system of free spins in
an external field:
\begin{equation}
H = \sum_i n_i\;.
\label{eq.H}
\end{equation}
From the thermodynamic point of view, this model is trivial and its
landscape is structureless. Being irreducible in the phase space, the 
equilibrium properties correspond to those of a paramagnet in a field: 
there is no phase transition and the concentration of
up spins, $c=1/(1+e^{1/T})$, is a decreasing function of $T$. Thus, since 
up spins are those that facilitate the dynamics, it becomes slower as
$T$ decreases. In other words,
the introduction of constraints in the  dynamics,
independent of the Hamiltonian, forces the system to evolve through a subset of paths
in space-time which becomes increasingly limited as the temperature is lowered. The 
system needs to bypass energy barriers in order to relax and this
induces an effective roughness in the landscape. Strong
entropic effects are introduced exclusively by the dynamics, while it is important to
note that almost all configurations are allowed, 
detailed balance is fulfilled, and ergodicity
guaranteed \cite{CrRiRoSe00}. Both equilibrium and non-equilibrium dynamics of these
models have been extensively studied \cite{FrAn84,FrAn85,JaEi91,EiJa93,CrRiRoSe00,SoEv03}
(for a recent review see \cite{RiSo03}). Recently, a real space-time interpretation of 
the dynamics of glass formation has been put forward \cite{GaCh02,BeGa03,Be03} based on
observations on these kind of models. In this view, the up
spins are interpreted as regions of enhanced mobility while down spins are
nearly frozen regions. These regions are coarse grained both in time and space such
that no interaction is left and we get the effective one body Hamiltonian,
eq. \ref{eq.H}. These mobile regions, or defects, separate different domains (as
defined in \cite{SoEv03}) and the dynamics is described 
in terms of creation and annihilation
of defects, which induce a coarsening of domains. At any temperature there is a typical
lengthscale of the domains $l(T)=1/c(T)$, and the glass transition occurs at 
$T=0$, where this lengthscale diverges.

\subsection{The Symmetrically Constrained Ising Chain}
\label{SCIC}

We consider a $d=1$ chain of $N$ spins with periodic boundary 
conditions. A spin can flip according to the following rules:
\begin{equation}
\left.
	   \begin{array}{l}
	   1 \to 0 ~~ with~probability~1\\
	   0 \to 1 ~~ with~prob.~~ \exp{\left(-\frac{1}{T}\right)}
	   \end{array} \right\}
\begin{array}{l}
i\!f\!f~{\rm .ANY.}~of \\
the~two~nearest\,neighbors~is~up
\end{array} \nonumber
\end{equation}

This model presents several characteristic relaxation times, all of an
Arrhenius form, the slower one (the equilibration time) growing with temperature 
as $\tau = e^{3/T}$.
In this sense the SCIC corresponds to a {\em strong glass}.

\begin{figure}[ht]
\includegraphics[width=5.5cm,height=6.8cm,angle=270]{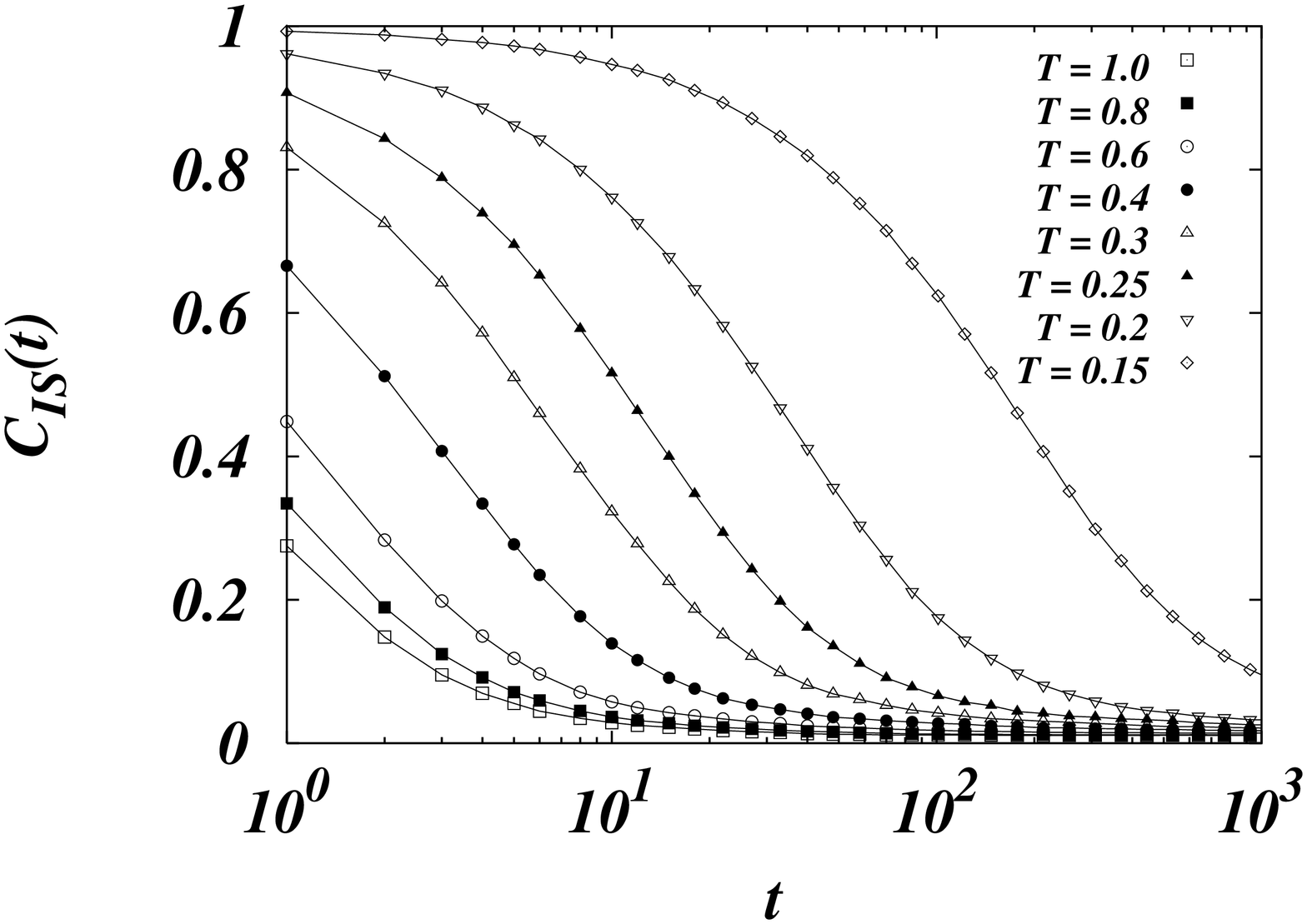}
\includegraphics[width=5.5cm,height=6.8cm,angle=270]{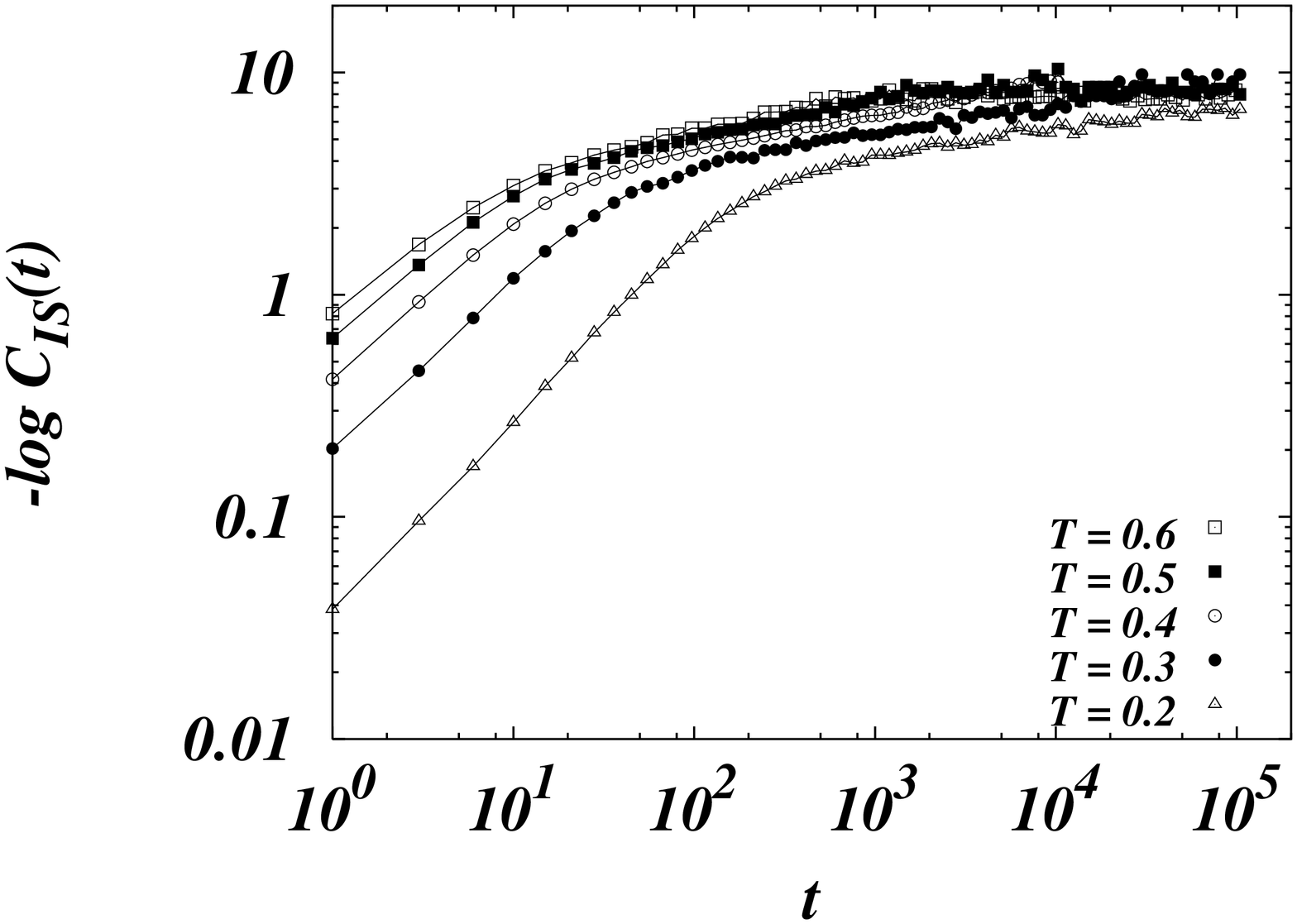}
\caption{\label{SCICfig}
(a) Temperature dependence of $\cis$ for the SCIC.
From left to right $T=1.0, 0.8, 0.6, 0.4, 0.3, 0.25, 0.2$ and $0.15$.
(b) $-\log\cis$.
Plotted in this way a stretched exponential
$exp[ -(t/\tau)^\beta$ ] is a straight line with slope $\beta$.}
\end{figure}

Following \cite{BeGa03} we have used a temperature dependent size in our simulations 
$L=4l$, where $l(T)=1/c(T)$ is the typical size of domains at temperature $T$. 
Initial states with all spins down are discarded. In figure \ref{SCICfig}a and b we 
show $\cis$ for the SCIC. 
In figure \ref{SCICfig}b we see the two time regimes observed also in the LJBM.
Nevertheless the dynamics of IS in the SCIM at low temperatures is essentially that of
a set of independent random walkers, i.e. a diffusional dynamics, and consequently it is
faster than the LJBM dynamics. 

\begin{figure}[ht]
\includegraphics[width=5.5cm,height=7.8cm,angle=270]{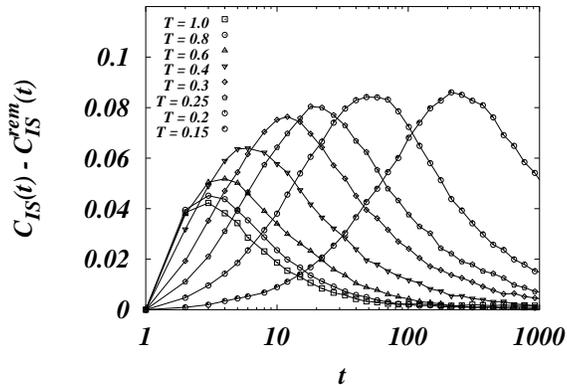}
\caption{\label{dif_SCIC} The difference
$\cis-\cisrem$ for the SCIC for different temperatures. }
\end{figure}

In figure \ref{dif_SCIC} we show $\cis-\cisrem$ for this model. Besides the
growing characteristic time scale with decreasing temperature we observe a tendency to
saturation in the value of the peak probability, which implies that the degree of confinement 
becomes nearly independent of $T$ at low temperatures. One can rationalize this
behavior, clearly different from what is observed both in the LJBM and in the trap model,
by analyzing the microscopic dynamics of defects. The mechanism of relaxation is the
coarsening of domains. Only with the annihilation of defects the energy can be reduced.
From an initial inherent structure, the SCIC is able to coalesce two domains into one by
diffusing the defects separating them. In order to diffuse a single defect from an IS it
is necessary to excite a neighbor site creating a new defect, what costs an energy
equal to unity. Once two defects are together then it is possible to relax this structure in
both directions and in this way allow the defects to diffuse. This climbing of an
energy hill of unit height and then relaxing again can go on, until the diffusing defect 
meets another one. Then one of them is annihilated and the collapse
of two domains happens. We can see that, although the typical size of domains grows with
decreasing temperature, the typical cost in order to collapse domains is always the same,
as it is only necessary to go up and down by steps of cost 1 in order to diffuse and
eventually relax. In other words, the confinement does not grow with decreasing 
temperature, because energy barriers are temperature independent. This produces a saturation
in the typical probability that the system returns to the initial inherent structure at
low temperatures. Clearly this behavior is different from the one observed in the LJBM and
help us to understand the process of relaxation in that model too. One could guess that energy
barriers do grow with decreasing temperature in the LJBM, inducing a growing confinement.
The temperature dependence of energy barriers is explicitly realized in the asymmetrically
constrained Ising chain, and its consequences are described in the next section.

\subsection{The Asymmetrically Constrained Ising Chain}
\label{ACIC}

The dynamics of the ACIC is defined as:

\begin{equation}
\left.
	   \begin{array}{l}
	   1 \to 0 ~~ with~probability~1\\
	   0 \to 1 ~~ with~prob.~~ \exp{\left(-\frac{1}{T}\right)}
	   \end{array} \right\}
\begin{array}{l}
i\!f\!f~{\rm .THE~LEFT.}\\
neighbor~is~up
\end{array} \nonumber
\end{equation}

At low temperatures the ACIC has a dominant relaxation time $\tau = e^{1/T^2 \ln 2}$
 \cite{SoEv03}. This behavior is super-Arrhenius corresponding to a
{\em fragile glass}. From this point of view it should be similar to the LJBM which is
also considered to be a model of a fragile glass former.

\begin{figure}[ht]
\includegraphics[width=5.5cm,height=6.8cm,angle=270]{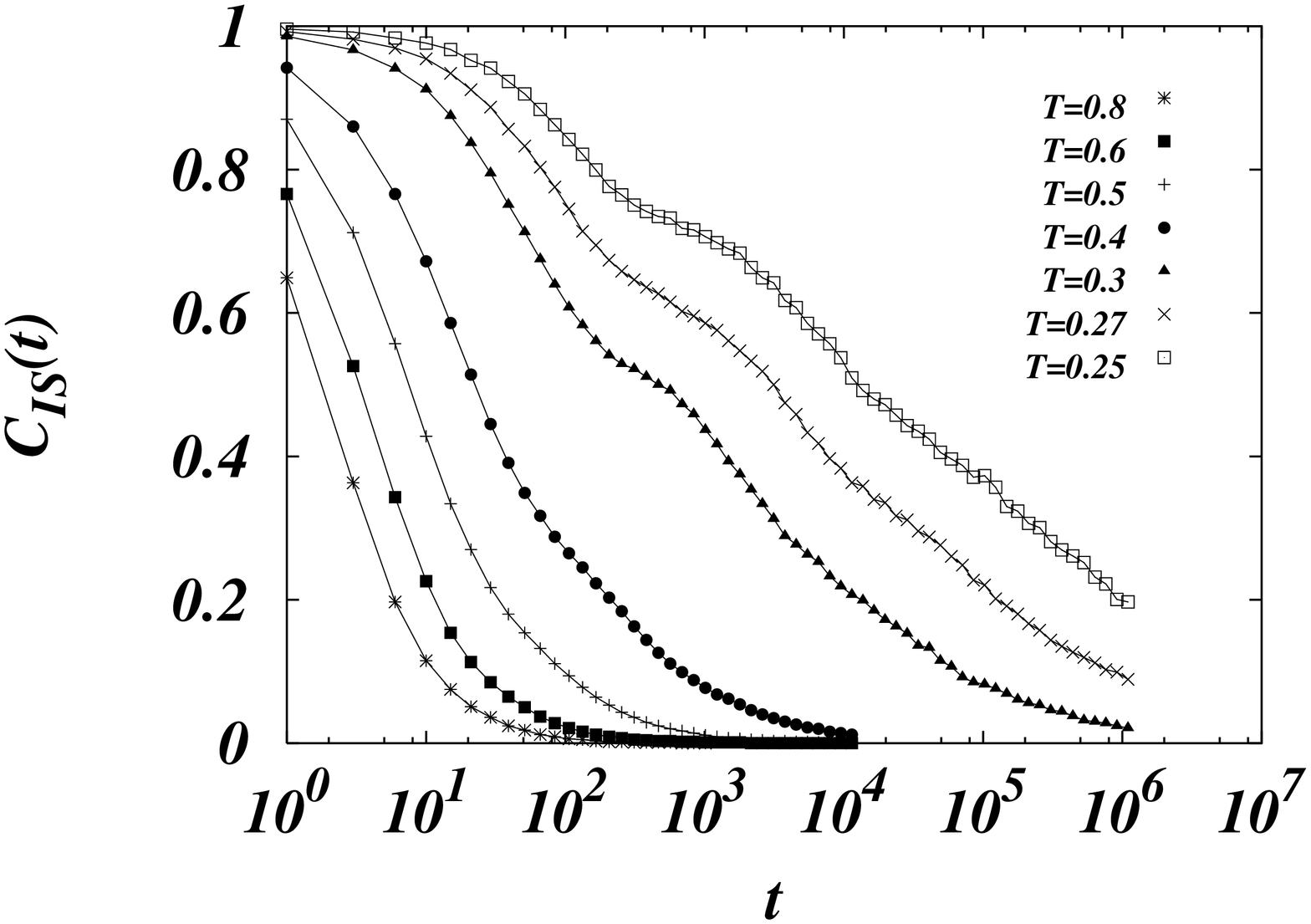}
\includegraphics[width=5.5cm,height=6.8cm,angle=270]{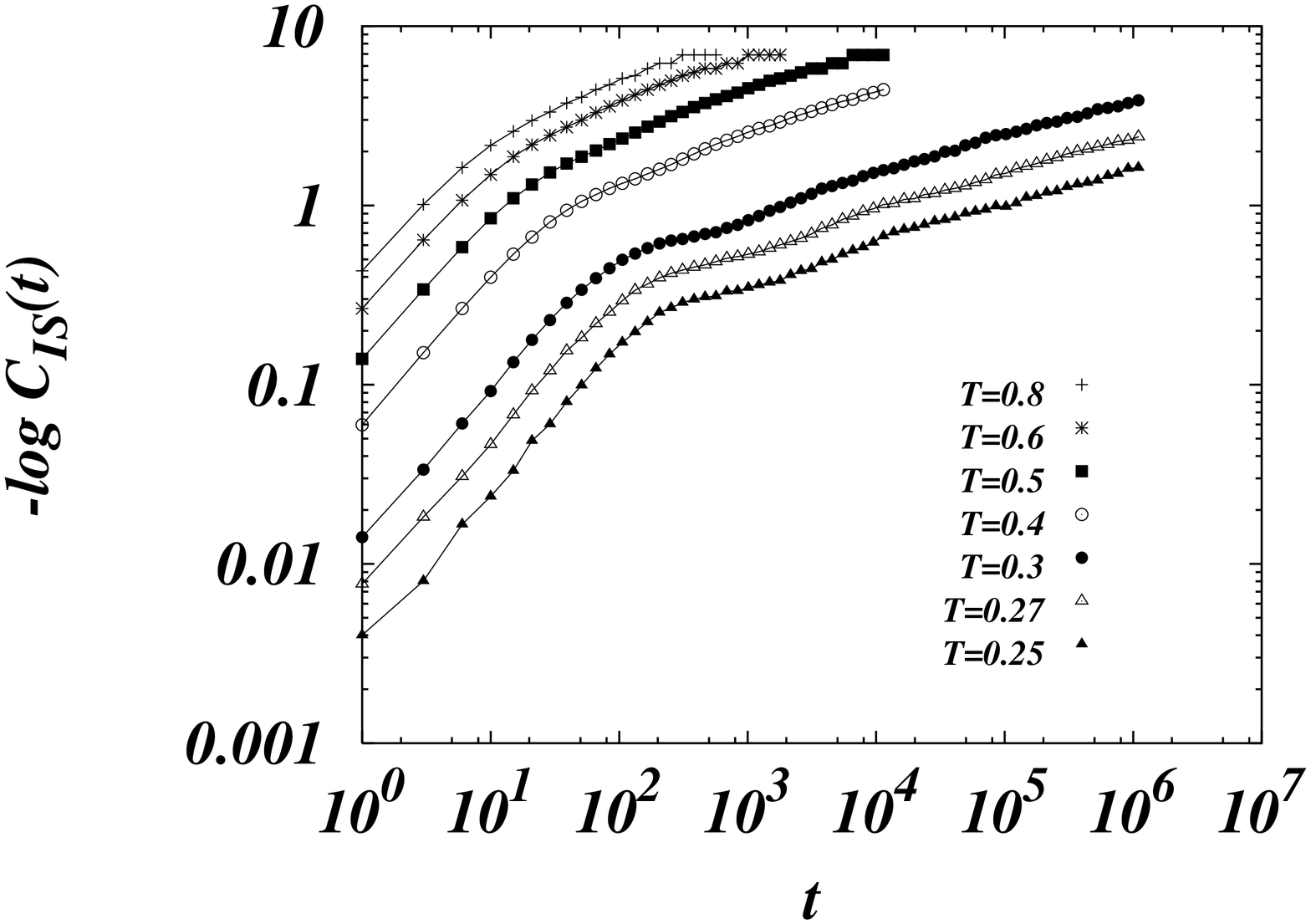}
\caption{\label{ACICfig}
(a) Temperature dependence of $\cis$ for the ACIC.
From left to right $T=0.8, 0.6, 0.5, 0.4, 0.3, 0.27$ and $0.25$.
(b) $-\log\cis$. Plotted in this way a stretched exponential
$\exp[ -(t/\tau)^\beta]$ is a straight line with slope $\beta$.}
\end{figure}

In this case the sizes simulated at each temperature were $L=8l$ \cite{BeGa03}. 
In figure \ref{ACICfig}a and b the results for the $\cis$ for the ACIC are shown. 
By comparing figures \ref{LJfig}, \ref{SCICfig} and \ref{ACICfig} one immediately 
recognizes a similar behavior
between the LJBM and the ACIC. In this case a short and a long time regimes
are again observed in the correlations between IS. A more detailed inspection shows that
stretched exponential relaxations are also present, although in this case the
stretching exponent is nearly independent of temperature, being close to $\beta\approx 0.2$
for the three smaller temperatures simulated. As expected, the relaxation times of 
the stretched
exponential behave as $\tau=e^{a/T^2}$, although the coefficient $a\approx 0.87$
from a fit of the three lower temperatures is nearly a half of the value $1/\ln 2$
corresponding to the largest relaxation times for the model.

\begin{figure}[ht]
\includegraphics[width=5.5cm,height=7.8cm,angle=270]{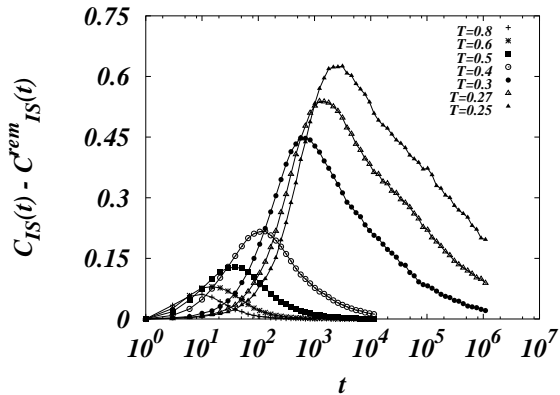}
\caption{\label{dif_ACIC} The difference
$\cis-\cisrem$ for the ACIC for different temperatures. }
\end{figure}

In figure \ref{dif_ACIC} we show $\cis-\cisrem$, that presents a remarkable
similarity with the corresponding curves for the LJBM, figure \ref{dif_LJ}. The typical
probability to come back to a particular inherent structure grows with decreasing
temperature, indicating a strong confinement in configuration space for the ACIC at low
temperatures. In this case this is a consequence of growing energy barriers with
decreasing temperature. The dynamics of relaxation and coarsening of domains in this
model is very different from that of the SCIC. Due to the asymmetry in the constraint,
there is a directionality in the dynamics of defects. In order to collapse two defects
initially separated by a sequence of down spins the only possibility is to propagate the
defect on the left until it reaches the spin in the right. But this propagation cannot
proceed by simple diffusion as in the SCIC. New excitations have to be created and
anchored until the defect on the right is reached. Then the reverse path can be taken
until the domains are completely coalesced. The complete process has an energy cost $h(l)$
which depends on the length of the domain $l$, given by \cite{SoEv03}:

\begin{equation}
h(l) = n + 1 \hspace{2cm} {\rm for}\,\,\,\,2^{n-1}<l\leq2^n.
\end{equation}

Consequently energy barriers in the ACIC are temperature dependent through the dependence
of $l(T)$ and grow with decreasing temperature. This produces confinement in configuration
space which makes the peak probability to return to an IS to grow as $T$ is lowered,
similar to what is observed in the LJBM. 

\section{Conclusions}
\label{final}

We did a comparative study of several different glass forming models 
from the point of view of the dynamics of inherent structures, by measuring appropriate
temporal correlations between these IS as the temperature is lowered. These
correlation functions do show signatures of the complex structure of the configuration
space, as the presence of confinement at low temperatures.
The models we considered have very different origins, the Lennard-Jones binary mixture 
which is a Hamiltonian molecular model, the trap model which is a phenomenological model 
with purely activated dynamics, and two kinetically constrained models which are discrete 
lattice models with Monte Carlo dynamics. Regarding the properties considered here, all 
four models show very different behavior, the trap model being the less interesting one. 
By construction it may be thought as having a flat landscape filled with holes of random 
depths. The correlations show that lowering the temperature the probability for the system 
to return to an IS diminishes since there is no mechanism forcing it to stay in a region 
except the residence time inside the traps. The model that is next in complexity is perhaps the
symmetrically constrained Ising chain, a model of a strong glass. This model behaves 
differently from the trap model in that it shows a growing probability to return
to an IS at lower temperatures, an evidence of the non trivial character of the configuration
space. Nevertheless this effect is rather weak as evidenced by an early saturation of 
the peak
probability to return. This is a consequence of the temperature independence of the energy 
barriers. The system relaxes by diffusion of defects with a constant cost. In
the last two explored models, the Lennard-Jones binary mixture and the asymmetrically
constrained Ising chain the behavior is much more interesting. In both models the
peak probability to return to an IS steadily grows with decreasing temperature
signalling a strong confinement on both systems. The mechanism behind this behavior can
be understood in terms of energy barriers in the ACIC. The typical barriers that
the system has to cross in order to coalesce two domains grow with decreasing $T$,
differently from what happens in the SCIC. This points to a common origin of relaxations
in the ACIC and in the LJBM. It is clear that although the ACIC posses a trivial
equilibrium measure, the constraint imposed on the dynamics makes it evolve in an
effective non trivial landscape. In the LJBM this effect is more fundamental, it comes
directly from the microscopic interactions which produce a higly non trivial energy
landscape which is in turn the ultimate origin of the complex glassy dynamics in this
system. In summary, we found that for models where the evolution follows by 
cooperative rearrangements, there is an increase in the returning probability, that
is, a stronger confinement. We also expect that this is a rather general property,
and should be valid in other systems depending on the cooperativeness of the
dynamics.

This work was supported in part by CONICET and {\it Fundaci\'on Antorchas}, Argentina
and by CNPq, Brazil. We also acknowledge useful discussions with Juan P. Garrahan.


\end{document}